\documentclass[twocolumn]{article}
\usepackage{spconf,amsmath,graphicx}
\usepackage{booktabs}
\usepackage{cite}
\usepackage{stfloats}
\usepackage{bm}
\usepackage{multirow}
\usepackage{subfiles}
\usepackage{subfigure}
\usepackage{amsfonts,amssymb}
\usepackage{lipsum}
\usepackage{microtype}

\title{Frequency bin-wise single channel speech presence probability estimation using multiple DNNs}
\name{Shuai Tao, Himavanth Reddy, Jesper Rindom Jensen, Mads Græsbøll Christensen}
\address{Audio Analysis Lab, CREATE, Aalborg University, Aalborg, Denmark\\
stao@create.aau.dk, hire@create.aau.dk, jrj@create.aau.dk, mgc@create.aau.dk}

\begin{document}
\ninept
\maketitle

\newcommand\blfootnote[1]{%
\begingroup
\renewcommand\thefootnote{}\footnote{#1}%
\addtocounter{footnote}{-1}%
\endgroup
}
\begin{abstract}
In this work, we propose a frequency bin-wise method to estimate the single-channel speech presence probability (SPP) with multiple deep neural networks (DNNs) in the short-time Fourier transform domain. Since all frequency bins are typically considered simultaneously as input features for conventional DNN-based SPP estimators, high model complexity is inevitable. To reduce the model complexity and the requirements on the training data, we take a single frequency bin and some of its neighboring frequency bins into account to train separate gate recurrent units. In addition, the noisy speech and the $a$ $posteriori$ probability SPP representation are used to train our model. The experiments were performed on the Deep Noise Suppression challenge dataset. The experimental results show that the speech detection accuracy can be improved when we employ the frequency bin-wise model. Finally, we also demonstrate that our proposed method outperforms most of the state-of-the-art SPP estimation methods in terms of speech detection accuracy and model complexity.

\end{abstract}
\begin{keywords}
frequency bin-wise, speech presence probability, $a$ $posteriori$ probability, gated recurrent units
\end{keywords}
\section{Introduction}
\label{sec:intro}

Noise estimation is one of the key components to realize single-channel and multi-channel speech enhancement, most of which rely on the speech presence probability (SPP) to update the noise statistics \cite{kim2022improved, roy2022robustness, zhao2020model}. Available noise power spectral density (PSD) estimators also make use of the SPP to decide when to update the noise PSD \cite{cohen2003noise, gerkmann2011unbiased, souden2011integrated}. Compared to voice activity detectors (VAD), SPP is a soft-decision approach that depends on the correlation of inter-bands and inter-frames \cite{momeni2014single}. Accurate SPP estimation can greatly improve the effectiveness of speech enhancement \cite{souden2009gaussian, rangachari2006noise}.

In the short time-frequency transform (STFT) domain, some conventional statistical signal processing methods commonly assume that the spectral coefficients of speech and noise are independent and follow the complex Gaussian distribution \cite{mcaulay1980speech, ephraim1984speech}. Therefore, the SPP can be derived from the $a$ $posteriori$ probability of the time-frequency (T-F) bins of the noisy speech. According to this assumption, \cite{cohen2003noise} applied the minima values of a smoothed periodogram to estimate the SPP which enables the SPP estimation to be more robust under the effect of non-stationary noise. In \cite{gerkmann2011unbiased}, to achieve a highly accurate SPP estimate with low latency and computational complexity, an optimal fixed $a$ $priori$ SNR was used to guarantee the $a$ $posteriori$ SPP to be close to zero when speech is absent. In addition, \cite{momeni2014single} takes the correlation of inter-band and inter-frame into account when designing a general SPP estimator.

Recently, deep neural networks (DNNs) have been proven to be effective at processing non-stationary noise, and many novel DNN-based approaches have been proposed to estimate SPP accurately, which have been applied to speech enhancement and speech recognition successfully \cite{tu2019speech, tammen2020dnn, zhang2020deepmmse}. In these methods, recurrent neural networks (RNNs) \cite{pearlmutter1995gradient} are commonly used to acquire information from neighboring frames since the frames contain temporal information which can improve the accuracy of SPP estimation. In \cite{zhang2020deepmmse}, a bidirectional long short-term memory (BLSTM) was trained by the input features of multi-time frames with all frequency bins to estimate the SPP. In \cite{tu2019speech}, considering the ideal ratio mask (IRM) \cite{wang2014training} ranges from 0 to 1 at each T-F bin, they selected different DNN models, such as LSTM, BLSTM, gate recurrent units (GRUs), and bidirectional GRU (BGRU) to estimate the IRM and approximate the SPP. However, the problem that arises here is that as the complexity of the model goes up and more training data is applied to the model, more powerful hardware is required to train the models.

Inspired by conventional SPP estimation methods, our model estimates the SPP based on the correlation of several neighboring T-F bins in contrast to the typical DNN-based SPP estimation approach where all frequency bins are regarded as the input features. This allows us to use DNNs on a one-to-one basis with frequency bins therefore vastly reducing the number of parameters in the model and the amount of computations taking place. In this work, we thus propose a frequency bin-wise SPP estimation model in the STFT domain that relies on using multiple DNNs to estimate the SPP. For our proposed model architecture, the GRU module is used to extract time and frequency information from each frequency bin and several of its neighbors. Additionally, since IRM-based SPP estimation methods may misclassify the T-F bins dominated by non-speech and noise \cite{subramanian2018student, chen2018building, tu2019speech}, we choose the $a$ $posteriori$ probability to represent the SPP in the STFT domain. 

The work is organized as follows. In Section 2, the problem of frequency bin-wise single channel SPP estimation is formulated. In Section 3, the SPP estimation model with multiple DNNs is designed. In Section 4 and Section 5, the experimental procedures and results are provided, respectively. Finally, Section 6 presents the conclusion. The work can be found on GitHub$\footnote{https://github.com/Shuaitaoaau/SPP}$.

\section{Frequency Bin-Wise SPP Estimation}
\label{sec:Theory}
\subsection{Signal Modeling}
For the single channel speech signal $x(n)$, we assume that it is corrupted by the additive noise $d(n)$. That is, in the STFT domain, we can obtain the noisy speech $y(n)$ representation as follows:
\begin{equation}
  Y(k,l)=X(k,l)+D(k,l),
\end{equation}
where $k\in \{0,...,K-1\}$ denotes the frequency bin index and $K$ is the number of frequency bins, $l\in \{0,...,L-1\}$ denotes the time frame index and $L$ is the number of time frames. With the assumption of a zero-mean complex Gaussian distribution and independence for $X$ and $D$, we have
\begin{equation}
\begin{split}
    \bm{\phi}_Y(k,l)&=E[|Y(k,l)|^2]\\
    &= \bm{\phi}_X(k,l)+\bm{\phi}_D(k,l),
\end{split}
\end{equation}
where $E[\cdot]$ is the statistical expectation operator, ${\phi}_X(k,l)=E[|X(k,l)|^2]$ and ${\phi}_D(k,l)=E[|D(k,l)|^2]$. The PSD of the clean and the noisy speech can be represented by $\phi_X(k,l)$ and $\phi_D(k,l)$, respectively. In the STFT domain, there exists a correlation between the neighboring T-F bins \cite{momeni2014single}. Therefore, the SPP estimate can be improved using the correlation.

The first step in creating our input signal vector is to obtain a vector corresponding to each individual frequency bin,
\begin{equation}
    \bm{\varphi}_Y(k)=[\phi_Y(k,0),...,\phi_Y(k,l),...\phi_Y(k,L-1)]^T.
\end{equation}
Each frequency bin vector contains $L$ consecutive time frames, which contain relevant contextual information for the estimation of the SPP. Since RNNs are effective at processing temporal information \cite{wang2021tstnn, pandey2022self}, we employ RNNs in this work to extract time correlations from the neighboring time frames.

To improve the SPP estimation accuracy, we take a few neighboring frequency bin vectors into consideration to extract frequency correlations from the input signal matrix. Therefore, the input signal matrix $\bm{\Phi}_Y(k)$ can be obtained as
\begin{equation}
    \bm{\Phi}_Y(k)=[\bm{\varphi}_Y(k-I),...,\bm{\varphi}_Y(k),...,\bm{\varphi}_Y(k+I)]^T,
\end{equation}
where $I$ is the number of neighboring frequency bin vectors.

Now, the time correlation and frequency correlation of neighboring time-frequency bins can be extracted according to the input signal matrix $\bm{\Phi}_Y(k)$. In this work, the SPP is represented by the \textit{a} \textit{posteriori} probability \cite{gerkmann2011unbiased}, and the DNN is used to estimate the SPP from the noisy observation.

Since the typical DNN-based approach takes all the frequency bins into account to estimate the SPP, the model complexity may be increased. In this section, we, therefore, design multiple specific DNNs to estimate the frequency bin-wise SPP. Additionally, since the $a$ $posteriori$ probability is derived by the correlation of neighboring T-F bins, the $a$ $posteriori$ probability SPP representation of the clean speech and the noisy speech PSD are used as the training data pairs to train our model.

\begin{figure}
    \centering
    \subfigure[The typical DNN-based SPP estimation model training strategy]{
    \includegraphics[width=8.5cm]{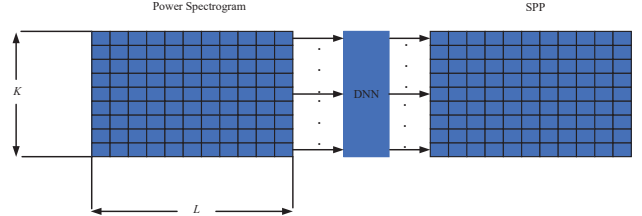}
    }
    \quad
    \subfigure[Frequency bin-wise SPP estimation model training strategy]{
    \includegraphics[width=8.5cm]{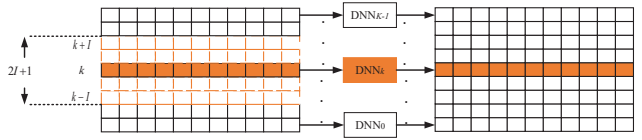}
    }
    \caption{Typical DNN-based model training strategy vs our proposed method. (a) Typical DNN-based SPP estimation model (with all frequency bins), and (b) Proposed frequency bin-wise
    SPP estimation model, a frequency bin along with $2I$ neighboring frequency bins are treated as the input features.}
\end{figure}

\subsection{SPP Estimation Model and Loss Function}
To extract the time and frequency correlation of the consecutive T-F bins in the input signal matrix $\bm{\Phi}_Y(k)$ from the observed noisy PSD $\bm{\phi}_Y(k,l)$, we set $K$ specific DNNs as the regression module. As mentioned in (4), the coefficient of the $k$'th input signal matrix can be used to train the $k$'th DNN for the SPP estimate in the $k$'th frequency bin.

First, to train the DNN model, we choose the log-power periodogram as the input feature \cite{xu2014regression, mirsamadi2016causal}. Therefore, the input features of each individual DNN are obtained from the log input signal matrix $\bm{\Phi}_Y(k)$. It can be expressed as
\begin{equation}
\begin{split}
     \bm{\Phi'}_Y(k)&=\log(\bm{\Phi}_Y(k)),
\end{split}
\end{equation}
where $\bm{\Phi'}_Y(k)$ is the input feature for the $k$'th DNN. Also, during training, we have

\begin{equation}
    \widehat{\text{SPP}}_Y(k)=F^\theta_k(\bm{\Phi'}_{Y}(k)),
\end{equation}
where $\widehat{\text{spp}}_{Y}(k) = [\widehat{\text{SPP}}_{Y}(k,0),...,\widehat{\text{SPP}}_{Y}(k,l),...,\widehat{\text{SPP}}_{Y}(k,L-1) ] ^T$ is the SPP estimate of the $k$'th input features, $F^\theta_k$ is the $k$'th DNN with the parameter $\theta$.
To update the DNN parameters, the loss between the target and the estimated SPP is calculated by mean-squared error (MSE), i.e.,
\begin{equation}
  \begin{split}
    L_{MSE}= \frac{1}{L}\sum_{l=0}^{L-1}(\text{SPP}_{Y}(k)-\widehat{\text{SPP}}_{Y}(k))^2,
  \end{split}
\end{equation}
where $\text{SPP}_{Y}(k)=[\text{SPP}_{Y}(k,0),...,\text{SPP}_{Y}(k,l),...,\text{SPP}_{Y}(k,L-1) ] ^T$ is the target function. In this work, the $a$ $posteriori$ probability is regarded as the SPP representation, therefore $\text{SPP}_{Y}(k,l)$ can be represented by
\begin{equation}
  \text{SPP}_{Y}(k,l)
        =\left(1+\frac{p(\mathcal{H}_0)}{p(\mathcal{H}_1)} \left(1+\xi_{\mathcal{H}_1} \right) e^{-\frac{|Y|^2}{\bm{\phi}_{D}}\frac{\xi_{\mathcal{H}_1}}{1+\xi_{\mathcal{H}_1}}}\right)^{-1}
\end{equation}
where $p(\mathcal{H}_{0})$ and $p(\mathcal{H}_{1})$
 denote $a$ $priori$ speech absence and presence probability, $\xi_{\mathcal{H}_{1}}$ is the $a$ $priori$ SNR during speech presence \cite{gerkmann2011unbiased}.
 
\subsection{Model Architecture}
In this work, since a GRU can outperform an LSTM both in terms of convergence in CPU time, and in terms of parameter updates and generalization \cite{chung2014empirical}, we choose GRUs to design the SPP estimation model. The model training strategy is shown in Fig. 1 and the DNN model is trained by the input features of the logarithmic power spectral T-F bins.

The training strategy of the typical DNN-based SPP estimation model in Fig. 1(a) shows that a GRU module is trained using $K$ frequency bins (all frequency bins) and $L$ consecutive time frames. The typical DNN-based model input size is $K$ and, in this work, the size of the hidden layer is the same as the size of the input layer. The proposed training strategy of the frequency bin-wise SPP estimation model is shown in Fig. 1(b). When $I$ neighboring frequency bins are introduced to estimate the SPP of a single frequency bin, the input size is $2I+1$, and one hidden layer is set. The output of each hidden layer state is regarded as the value of the SPP estimate at the current time. Finally, to restrict the output range of the DNN to [0, 1], the output layer is the activation function $Softplus$ with a fixed parameter $\beta$.

\section{Experimental Settings}
In this work, the sub-band DNS dataset is used to train our designed model. During testing, 200 noisy utterances (1.1 hours) and 1800 noisy utterances (1 hour) were collected from the DNS dataset \cite{dubey2022icassp}, and the TIMIT dataset \cite{garofolo1993darpa}, respectively. Each clean utterance is corrupted by a random noise utterance selected from the noise dataset, each noisy utterance SNR ranging from -5dB to 25 dB. The noise data includes 150 different types of noise taken from Audioset \cite{gemmeke2017audio} Freesound \cite{font2013freesound} and Demand datasets \cite{thiemann2013diverse}.

The receiver operating characteristic (ROC) \cite{cook2007use} curve is used to evaluate the SPP estimation method performance and the false-alarm probability $P_\text{fa}=0.05$ given in \cite{momeni2014single} is used to calculate the speech detection probability, $P_\text{d}$. Additionally, we apply the area under curve (AUC) metric which is derived from ROC and ranges between [0, 1] to represent overall performance. We also adopt the adaptive threshold set to -60 dB below the maximum instantaneous power across all TF bins shown in \cite{momeni2014single} to distinguish the speech and non-speech bins across all T-F bins of clean speech.

The sampling rate of all utterances is 16 kHz. Hann window is applied to STFT analysis and the length of the time window for STFT is 16 ms and the hop length is 8 ms. We use the mean and standard derivation to normalize the dataset. During training, the Adam optimizer \cite{kingma2014adam} is utilized to optimize the neural network parameters. The learning rate is set to 0.001. Weight decay is set to 0.00001 to prevent overfitting. The parameter will be updated at the 50th and 100th epochs for the implemented DNN models. Pytorch is used to implement the frequency bin-wise SPP estimation model and the reference DNN-based model.

\section{Results and Discussion}
In this section, to prove the effectiveness of our method, a comparison is shown between a typical DNN-based model and our proposed method using ROC curves. Moreover, some numerical results are provided to evaluate the accuracy of the SPP estimators and the model complexity, respectively.

\subsection{Examination of ROC Curves}

\begin{figure}[htb]
    \centering
    \centerline{\includegraphics[width=7.5cm, height=4cm]{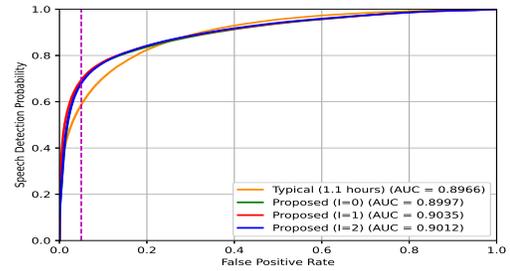}}
    \caption{ROC curves comparison of the typical DNN-based model and the frequency bin-wise model. Both models are trained with the same amount of training data (1.1 hours). The vertical dotted line indicates the false-alarm probability $P_\text{fa}=0.05$. Input SNR = 11 dB.}
    \label{ROC Curve 1}
\end{figure}

To investigate the performance of the proposed method, 200 training utterances (1.1 hours) are used to train our proposed frequency bin-wise model. In addition, 200 utterances (1.1 hours), 1000 utterances (5.5 hours), and 3000 utterances (16.6 hours) are used to train the typical DNN-based model, respectively. To investigate the effect of using neighboring frequency bins for the proposed method, we set $I = 0$ (no neighboring frequency bins), $I = 1$ (with 1 neighboring frequency bin), and $I=2$ (with two neighboring frequency bins) to train the frequency bin-wise model. Fig. \ref{ROC Curve 1} shows an example of SPP estimation results. A noisy utterance of length 20 seconds  and input SNR of 11 dB taken from the DNS dataset, is used for testing by the typical DNN-based SPP estimation model and the frequency bin-wise model.

\begin{figure}[htb]
    \centering
    \centerline{\includegraphics[width=7.5cm, height=4cm]{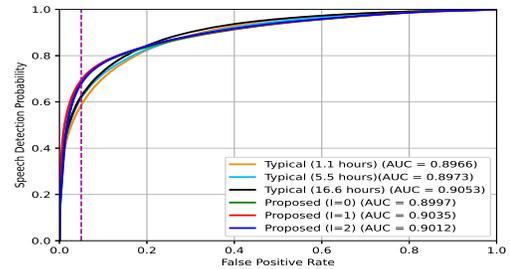}}
    \caption{ROC curves comparison of the typical DNN-based model and the frequency bin-wise model with an increase in training data for the typical DNN-based model. The vertical dotted line indicates the false-alarm probability $P_\text{fa}=0.05$. Input SNR = 11 dB.}
    \label{ROC Curve 2}
\end{figure}

From Fig. \ref{ROC Curve 1}, we can observe that the typical DNN-based method and the proposed frequency bin-wise method are able to estimate the SPP with similar accuracy. In addition, we also investigate the impact of the training data volume on SPP estimation accuracy for the typical DNN-based SPP estimation model. From Fig. \ref{ROC Curve 2}, we can find that when we increase training data from 1.1 hours to 5.5 hours and then to 16.6 hours for the typical DNN-based model, there is a gradual increase in AUC but still falls short of our proposed method in terms $P_\text{d}$.

\subsection{Numerical Results}
To evaluate the performance of the proposed method, the speech detection probability and the AUC are calculated from the ROC curves to represent the speech detection accuracy and the effectiveness of the SPP estimation method, respectively. In addition, we also investigate the effect of model complexity on SPP estimation accuracy. Inspired by \cite{vaswani2017attention} and \cite{devlin2018bert}, we compare our method with the state-of-the-art self-attention model and, in this work, 3 self-attention heads and 2 encoder layers are used to estimate the SPP. The self-attention model is trained in a typical way where all the frequency bins are treated as input features. During training, the frequency bin-wise SPP estimation model and the self-attention-based SPP estimation model are trained with 1.1 hours of training data pairs. The typical DNN-based model is trained with 1.1 and 16.6 hours of training data pairs, respectively. All training data pairs come from the DNS dataset.

\begin{table}[t]
  \centering
  \caption{Speech detection probability $P_\text{d}$ and AUC for different SPP estimation methods. Input SNR $\in$ [-5 dB, 25 dB].\\}
  \begin{tabular}{ccc}
  \toprule
   Methods & $P_\text{d}$ ($P_\text{fa}=0.05$) & AUC \\
  \midrule
  IMCRA \cite{cohen2003noise}& 0.1183 & 0.6504\\
  Unbiased \cite{gerkmann2011unbiased} & 0.3460 & 0.7348\\
  General \cite{momeni2014single}& 0.1132 & 0.6229\\
  Self-Attention \cite{vaswani2017attention} (1.1 hours) & 0.4617 & \textbf{0.8100}\\
  Typical DNN-based (1.1 hours) & 0.4509 & 0.7993\\
  Typical DNN-based (16.6 hours) & 0.4652 & 0.8012\\
  Proposed ($I = 0$) (1.1 hours) & 0.5012 & 0.7986 \\
  Proposed ($I = 1$) (1.1 hours) & \textbf{0.5038} & 0.8011\\
  Proposed ($I = 2$) (1.1 hours) & 0.4891 & 0.7988\\
  \bottomrule
\end{tabular}
\label{AUC Table}
\end{table}

In Table \ref{AUC Table}, we show how the proposed model compares to other conventional methods and a few DNN-based methods using $P_\text{d}$ and AUC as metrics. The results in Table \ref{AUC Table} are obtained from testing using the TIMIT dataset (1 hour). 

With 1.1 hours of training data, we can observe that the frequency bin-wise model AUC (0.7986) is lower than the typical DNN-based model and the self-attention-based model, it is still higher than IMCRA \cite{cohen2003noise} (0.6504), Unbiased MMSE \cite{gerkmann2011unbiased} (0.7348) and General SPP estimator \cite{momeni2014single} (0.6229). Especially, when we set $I=1$ and $I=2$, the sub-frequency bin-based model achieved higher AUCs of 0.8011 and 0.7988, respectively. For the speech detection accuracy, all the frequency bin-wise models achieved higher speech detection accuracy than other methods and when we take one neighboring frequency bin ($I=1$) into account the speech detection probability can reach 0.5038. 

According to the results, we can confirm that an increase in model complexity can improve the performance of DNN-based applications, and in this work, the SPP estimation accuracy can also be improved, which is consistent with the experimental results shown in \cite{rehr2021snr}. The reason is that the complex model can extract more global information than the simple model to estimate the SPP from all frequency bins. Additionally, a remarkable improvement in speech detection accuracy appears when we employ our proposed method to estimate the SPP, especially when we set $I=1$, the model performance and $P_\text{d}$ are improved.  The reason for the improved performance could be that the DNNs can extract specific contextual information for each frequency bin which is not possible when $I=0$ due to the lack of inclusion of its neighbors.

Finally, by comparing the AUC of different SPP estimation methods, we can observe that all DNN-based models can achieve higher performance of SPP estimation than the conventional methods. For DNN-based SPP estimation models, although all the presented models demonstrate similar performance, the speech detection accuracy is different. Therefore, it can be observed that more details can be detected by the bin-wise model leading to better detection accuracy.

\subsection{Computational Complexity}
To evaluate the complexity of the proposed model relative to its counterparts, we use the number of parameters and floating point operations (FLOPs) as the metrics. For our proposed frequency bin-wise model, the total parameters and FLOPs of all the models are used to represent computational complexity. We use the \textit{ptflops} $\footnote{https://pypi.org/project/ptflops/}$ python library to calculate the total parameters and FLOPs for our method and the reference DNN-based methods. Table 2 shows that our proposed method has fewer parameters and FLOPs than the other methods. The reason is that although we use multiple DNNs to estimate the SPP, each DNN has less input size than the typical DNN-based model. Furthermore, although we introduced the neighboring frequency bins to estimate the SPP in 4.2, from Table 2, we can also observe that the increase in computational complexity is minimal even with the inclusion of additional neighboring frequency bins.

\begin{table}[t]
  \centering
  \caption{Parameters and FLOPs comparison of the DNN-based SPP estimation model.\\}
  \begin{tabular}{ccc}
  \toprule
   Methods & Para & FLOPs (Mac) \\
  \midrule
  Self-Attention \cite{vaswani2017attention} & 867.12K &  85.6M\\
  Typical DNN-based & 100.62K & 13.1M\\
  Proposed ($I = 0$) & \textbf{1548} & \textbf{2451} \\
  Proposed ($I = 1$) & 2292 & 3188\\
  Proposed ($I = 2$) & 3024 & 3920 \\
  \bottomrule
\end{tabular}
\end{table}

From the above experimental results, we can confirm that although increasing the training data and using complex models can contribute to the improvement of the performance of the typical DNN-based SPP model, high computational complexity is inevitable. However, it can be observed that the proposed frequency bin-wise model not only shows an improvement in $P_\text{d}$ while maintaining similar performance in terms of the AUC but also reduces the computational complexity while using the same amount of training data.

\section{Conclusion}
In this work, we proposed an effective frequency bin-wise SPP estimation method that shows good performance with a limited amount of training data while also maintaining low model complexity. Experimental results show that in addition to reducing the model complexity, the frequency bin-wise model also shows better performance even in comparison with the typical DNN-based model that is trained with increasing amounts of training data. The experimental observations involving the inclusion of neighboring frequency bins show that there is an increase in speech detection accuracy as well as the AUC (compared to its counterpart that does not include any neighboring frequency bins) due to being exposed to local contextual information. Since multiple DNNs are employed to estimate the SPP in the STFT domain, the frequency bin-wise model's computational complexity is much lower than its DNN-based counterparts.



\bibliographystyle{IEEEtran}
\bibliography{strings}

\end{document}